\documentclass[aps,prc,reprint,amsmath,amssymb,nofootinbib,superscriptaddress,longbibliography,floatfix]{revtex4-2}
\usepackage{graphicx}
\usepackage{booktabs}
\graphicspath{{fig/}}

\newcommand{\MeV}{\,\mathrm{MeV}}
\newcommand{\fm}{\,\mathrm{fm}}
\newcommand{\Msun}{M_\odot}
\newcommand{\Mmax}{M_{\max}}
\newcommand{\nz}{n_0}
\newcommand{\UL}{U_\Lambda}
\newcommand{\US}{U_\Sigma}
\newcommand{\UX}{U_\Xi}
\newcommand{\UN}{U_n}
\newcommand{\Ls}{\Lambda_{\rm s}}
\newcommand{\ksh}{k_{\rm sh}}
\newcommand{\kbu}{k_{\rm bu}}
\newcommand{\muB}{\mu_B}
\newcommand{\mue}{\mu_e}
\newcommand{\mukin}{\mu_{\rm kin}}
\newcommand{\Nc}{N_c}
\newcommand{\nc}{n_{\rm c}}
\newcommand{\nq}{n_{\rm q}}
\newcommand{\kq}{k_{\rm q}}
\newcommand{\Phat}{\overline{\lambda}}

\begin{document}

\title{Hyperons in quarkyonic neutron-star matter with in-medium potentials:\\
threshold conditions and maximum mass}

\author{J.\,A.\ Nowak}
\thanks{ORCID: \href{https://orcid.org/0000-0001-8637-5433}{0000-0001-8637-5433}}
\email[Corresponding author: ]{j.nowak@lancaster.ac.uk}
\affiliation{School of Physics and Astronomy, Lancaster University,
Lancaster LA1 4YB, United Kingdom}
\date{\today}

\begin{abstract}
\begin{description}
\item[Background] Fujimoto, Kojo and McLerran (FKM) proposed that in quarkyonic matter the Pauli
principle for quarks delays the appearance of $\Lambda$ and $\Sigma^0$ hyperons and limits their
abundance, so that their softening is mild. They use vacuum masses, neutral baryons and
occupation probabilities of a prescribed form.
\item[Purpose] To establish which of these conclusions hold with single-particle potentials,
a nucleonic sector supporting two solar masses, and occupation probabilities
obtained by minimising the energy.
\item[Method] The model of FKM is supplemented by a nucleonic energy functional and
hyperon potentials, fixed at nuclear saturation density $\nz$ by hypernuclear data. Above
$\nz$ they follow a parametrisation or chiral effective field theory. The energy
is minimised over all occupation probabilities, globally at zero temperature for the $\Lambda$ in
neutron matter and by iteration at $T=1\MeV$ for the octet in $\beta$ equilibrium. An uncertain factor $\xi$ between 0 and 1
sets how strongly hyperons source the mean field.
\item[Results] In neutron matter no $\Lambda$ appears continuously at density $n$ if
$\UL(n)-\UN(n)>D(n)$, where $\UN$ is the neutron single-particle potential and $D$, known in closed form, depends
only on the baryon size scale $\Ls$. In a model with $\Ls=400\MeV$, as used by FKM, a maximum mass of
$2.1\,\Msun$ requires $\UN\approx+200\MeV$ at the centre, and with chiral $\Lambda$ potentials the $\Lambda$
appears near $2\,\nz$. For the $\Lambda$ alone, these potentials held constant above $3\,\nz$ lower the maximum mass by
$0.11$--$0.46\,\Msun$ over the range of $\xi$. The FKM occupation probabilities, used there beyond their
stated range, give $0.01\,\Msun$. Continued linearly, the potentials
give up to $0.46\,\Msun$ at $\xi=0$ and at most $0.07\,\Msun$ for $\xi\ge2/3$. With the octet,
$\xi_Y$ of 0.4, 1 or the light-quark fraction and the parametrised potentials without extra repulsion above $\nz$, the $\Xi^-$
appears first and the reduction is $0.07$--$0.08\,\Msun$. In two further models, with $\Ls=350$ and $300\MeV$, $\UN$ at the centre is
below $+80\MeV$ and the chiral potentials satisfy the condition at all densities. Their reduction is
below $0.01\,\Msun$ for $\xi\ge0.4$ and up to $0.14$ and $0.06\,\Msun$ for $\xi=0$.
\item[Conclusions] When the stiffness comes from nucleonic repulsion the condition is
similar to that of a hadronic model. The outcome is then decided by the potentials above $3\,\nz$
and by the hyperon--hyperon interaction, which observations do not determine. In all tested sets
of one schematic functional, chiral potentials held constant above $3\,\nz$ meet the condition only
when the neutron potential at the centre is below about $+80\MeV$. Such sets reach at most
$2.11\,\Msun$. In the two further models the modified threshold raises the required $\Lambda$
potential, and the small neutron potential lets the condition hold. With the linear extrapolation the condition is also met for $\Ls=400\MeV$, at
$2.01\,\Msun$ with an incompressibility of $243\MeV$. A first-order transition is not excluded.
The incompressibility and GW170817 tidal deformability of the three models
exceed the empirical ranges.
\end{description}
\end{abstract}

\maketitle

\section{Introduction}
\label{sec:intro}

Pulsars of two solar masses \cite{demorest2010,antoniadis2013,fonseca2021,miller2021}
require the equation of state (EoS) of dense matter to be stiff at several times the
nuclear saturation density $\nz=0.16\fm^{-3}$. In hadronic models hyperons appear at
two to three times $\nz$, soften the EoS and lower the maximum mass $\Mmax$ by
several tenths of a solar mass. This is the hyperon puzzle
\cite{chatterjee2016,tolos2020,gal2016,burgio2021,Vidana:2018bdi,Sedrakian:2022ata}.

The remedies proposed within hadronic physics act on the hyperon interaction at high
density. They include hyperon couplings to vector mesons beyond the SU(6) values
\cite{Weissenborn:2011ut,bednarek2012,Oertel:2014qza}, density-dependent couplings
\cite{Fortin:2017cvt,Tolos:2016hhl}, and repulsive hyperon--nucleon--nucleon forces
\cite{Lonardoni:2014bwa,Haidenbauer:2016vfq,Logoteta:2019utx,Gerstung:2020ktv}. In
each case the single-particle potential of the hyperon must become strongly repulsive
above $2\,\nz$. Hypernuclear data fix these potentials only at and below $\nz$
\cite{gal2016}.

Quarkyonic matter \cite{McLerran:2018hbz} offers a different mechanism. Baryons occupy
a shell in momentum space, and the low-momentum states are saturated by quarks.
Fujimoto, Kojo and McLerran (FKM) constructed an explicit model of this kind, the
IdylliQ model \cite{Fujimoto:2023mzy}, and applied it to strangeness
\cite{Fujimoto:2024doc}. A neutron contains two $d$ quarks and a $\Lambda$ contains one.
Once the $d$-quark states are saturated by neutrons, a low-momentum $\Lambda$ can be added
only if neutrons are removed. FKM find that the threshold for the onset of the $\Lambda$ moves
from the baryon chemical potential $\muB=M_\Lambda$ to
\begin{equation}
  \muB = 2M_\Lambda - M_N ,
  \label{eq:fkm}
\end{equation}
and that the occupation of the low-momentum hyperon states is limited to a fraction of order
$1/\Nc^3$, with $\Nc=3$ the number of colours. These hyperons then appear late and soften the
EoS little. FKM note that hyperons at high momenta are almost free of this limit. They state that
such hyperons raise the pressure as well as the energy density and do not soften the EoS much.
FKM call the effect a statistical repulsion between baryon species, induced by
the quark Pauli blocking constraint \cite{Fujimoto:2024doc,Kojo:2026bnm}.

In Ref.~\cite{Fujimoto:2024doc} the baryons have their vacuum masses. Neutrons are combined
with degenerate $\Lambda$ and $\Sigma^0$ hyperons, and the $\Xi^0$ is added. FKM expect that the
$\Xi^0$, which contains no $d$ quark, can cause significant softening at high density. The
occupation probabilities are piecewise constant, of a prescribed form. Without hyperons, we find
that these inputs give a maximum mass of $1.46\,\Msun$.
Here these restrictions are removed. The baryons feel single-particle potentials, and
the nucleonic sector is made to support $2\,\Msun$. Protons, leptons and the full octet
are included. The occupation probabilities are obtained by minimising the energy.

\section{Model}
\label{sec:model}

\subsection{Quark Pauli blocking constraints}
\label{sec:idylliq}

In the model of Refs.~\cite{Fujimoto:2023mzy,Fujimoto:2024doc} a baryon of momentum
$\mathbf{k}$ contributes quarks of momentum near $\mathbf{k}/\Nc$, spread over a momentum
scale $\Ls$ that characterises the size of the baryon. Let $f_X(k)$ be the occupation probability of baryon species $X$
and $B_q^X$ the fraction of its quarks that have flavour $q$. The occupation of
quark flavour $q$ is
\begin{equation}
  f_q(\mathbf{q}) = \Nc^3 \sum_X B_q^X \int\! d^3k\;
     K\!\left(\left|\mathbf{q}-\mathbf{k}/\Nc\right|\right) f_X(k),
  \label{eq:fq}
\end{equation}
with the kernel of FKM, which in the present normalisation reads
\begin{equation}
  K(p) = \frac{e^{-p/\Ls}}{4\pi \Nc^3 \Ls^2\, p} ,
  \label{eq:kernel}
\end{equation}
with $\int d^3k\,K(|\mathbf{q}-\mathbf{k}/\Nc|)=1$.
The Pauli principle for quarks gives the quark Pauli blocking constraint of FKM,
\begin{equation}
  f_q(\mathbf{q}) \le 1 \qquad (q=u,d,s;\ \text{all } \mathbf{q}).
  \label{eq:pauli}
\end{equation}
For a filled sphere of radius $k_F$, Eqs.~\eqref{eq:fq} and \eqref{eq:kernel} give
$f_q(0)=B_q^X\Nc^3[1-e^{-x}(1+x)]$ with $x=k_F/(\Nc\Ls)$, which is Eq.~(B3) of
Ref.~\cite{Fujimoto:2024doc} (Appendix~\ref{sec:appmath}).

In pure neutron matter $B_d^n=2/3$. The $d$-quark states saturate, $f_d(0)=1$, at
the neutron Fermi momentum $\kq$ and density $\nq$. We call $\nq$ the quark
saturation density; it is distinct from the nuclear saturation density $\nz$. Above
$\nq$ the neutron occupation is $1/(B_d^n\Nc^3)=1/18$ below a momentum $\kbu$ and
unity in a shell $\kbu<k<\ksh$ \cite{Fujimoto:2023mzy}. We write $\mukin(n)$ for
the derivative of the kinetic energy density with respect to the neutron density.
Below $\nq$ it is the Fermi energy $E_F$. Above $\nq$ it is given by Eq.~(B7) of
Ref.~\cite{Fujimoto:2024doc}, reproduced as Eq.~\eqref{eq:mukin}, and rises faster
with density than $E_F$. This rise makes the EoS stiff. The neutron chemical
potential is $\mu_n=\mukin+\UN$, with $\UN$ the neutron single-particle potential. We use
$M_N=939.565\MeV$, the neutron mass, and $M_\Lambda=1115.683\MeV$. For
$\Ls=400\MeV$, the value used by FKM, $\kq=453\MeV$ and $\nq=2.55\,\nz$. For symmetric matter the
same equations give $2.58\,\nz$, the value quoted by FKM.

FKM add $d_Y$ degenerate hyperon species with occupation $1/(d_YB_d^Y\Nc^3)$ below a momentum
$k_Y$ and remove the neutrons there. They introduce these piecewise-constant functions as an
ansatz. They state that it is valid when the quark states saturate before the hyperon threshold is
reached, and for $k_Y<\kbu$. Beyond this range neutrons of the shell
can convert to hyperons of momentum above $\kbu$. We use the same form for the $\Lambda$ alone,
$d_Y=1$, with occupation $1/9$. Where the equilibrium value of
$k_Y$ would exceed $\kbu$ we set $k_Y=\kbu$. This continuation is ours. We refer to the
functions with $d_Y=1$ and this continuation as the FKM occupation probabilities.

\subsection{Interaction energy}
\label{sec:int}

The energy density is
\begin{equation}
  \varepsilon = \sum_X \int\!\frac{2\,d^3k}{(2\pi)^3}\, f_X(k)\sqrt{k^2+M_X^2}
     + \varepsilon_{\rm int} + \varepsilon_{\rm lep},
  \label{eq:eps}
\end{equation}
where $\varepsilon_{\rm lep}$ is the energy density of free electrons and muons. For
the nucleons we use
\begin{equation}
\begin{aligned}
  W(\tilde n,\delta) &= \tilde n\left[a u + b u^{\sigma} + S\,u\,\delta^2\right] g(u),\\
  g(u) &= \left[1+(u/u_c)^3\right]^{-1},
\end{aligned}
  \label{eq:W}
\end{equation}
with $u=\tilde n/\nz$, the asymmetry $\delta=(n_n-n_p)/(n_n+n_p)$ and the
symmetry-energy coefficient $S=19\MeV$. The constants $a$ and $b$ are fixed by the
binding energy $-16\MeV$ and zero pressure of symmetric matter at $\nz$. The exponent
$\sigma$ and the cut-off $u_c$ control the repulsion above $\nz$. The factor $g$
removes the interaction at high density, where the baryon shell supplies the
pressure. Without hyperons $\tilde n=n_n+n_p$, and the nucleon potentials are
$U_{n,p}=\partial W/\partial n_{n,p}$. The functional is schematic and is not fitted
to nuclear data beyond the two conditions at $\nz$.

For the hyperons $Y$ we write
\begin{equation}
\begin{aligned}
  \varepsilon_{\rm int} &= W(\tilde n,\delta)
    + \sum_Y n_Y\left[U_Y(\tilde n,\delta) - \xi_Y\,\frac{\partial W}{\partial\tilde n}\right],\\
  \tilde n &= n_n+n_p+\sum_Y \xi_Y n_Y .
\end{aligned}
  \label{eq:eint}
\end{equation}
The single-particle potential of every species is the derivative
$U_X=\partial\varepsilon_{\rm int}/\partial n_X$ at the actual composition, so the EoS
obtained from this functional is thermodynamically consistent
(Appendix~\ref{sec:appmath}). For a single hyperon in nucleonic matter this derivative is
$U_Y(\tilde n,\delta)$, which is an input.

The potentials at $\nz$ in symmetric matter are taken from hypernuclear and
$\Sigma$-atom data \cite{gal2016,saha2004}: $\UL=-28\MeV$, $\US=+30\MeV$ and
$\UX=-14\MeV$. Above $\nz$ they are not known. We use two continuations. The first is
\begin{equation}
  U_Y(\tilde n,\delta) = R_Y\,U_{\rm iso}(\tilde n,\delta) + \kappa_Y\,u + c_Y\,(u^2-u),
  \label{eq:uy}
\end{equation}
where $U_{\rm iso}$ is the average $(U_n+U_p)/2$ of nucleonic matter with density
$\tilde n$ and asymmetry $\delta$, $R_Y$ is the fraction of light quarks in $Y$,
and $\kappa_Y$ is fixed by $U_Y(\nz)$. In the average the term $\pm2Su\delta g$ of the
symmetry energy cancels. The damping factor $g$ leaves a dependence on $\delta$:
in model~A, $U_{\rm iso}$ at $3\,\nz$ is $36\MeV$ lower in neutron matter than in
symmetric matter.
Equation~\eqref{eq:uy} is a parametrisation and is not derived from the quark
model. The coefficient $c_Y$ measures additional repulsion above $\nz$. For $c_Y<0$
the potential is unbounded below and the pressure decreases at high density, from
$4.5$--$6.25\,\nz$ for $c_\Lambda=-5$ and $-2.5\MeV$, so only $c_Y\ge0$ is used.

The second continuation, for the $\Lambda$ only, is the potential in neutron matter
from Brueckner calculations with $\Lambda N$ and $\Lambda NN$ forces of chiral effective field
theory (EFT) \cite{Gerstung:2020ktv}. Its uncertainty band, called the chiral-EFT band below, is $-1.1$ to $+5.3\MeV$ at $2\,\nz$ and $+70.0$ to
$+91.6\MeV$ at $3\,\nz$. We interpolate with a cubic through the origin, these values and
$-33.8\MeV$ at $\nz$. For both edges the cubic has a single minimum, of $-34\MeV$ near $0.9\,\nz$,
and is convex up to $3\,\nz$.
The calculation ends at $3.5\,\nz$. Reference~\cite{Gerstung:2020ktv} states that the potential
exceeds $100\MeV$ at $3.5\,\nz$ and, by extrapolation, $300\MeV$ at $5\,\nz$. Above $3\,\nz$ we
either keep the potential at its value at $3\,\nz$ or continue it linearly with its slope there.
These two extrapolations are called constant and linear. The linear one gives $110$--$142\MeV$ at
$3.5\,\nz$ and $232$--$294\MeV$ at $5\,\nz$. It agrees with that statement at $3.5\,\nz$ and lies
below it at $5\,\nz$, where the extrapolation of Ref.~\cite{Gerstung:2020ktv} is quadratic in the
density. The constant one is the less repulsive of the two. The neutron-matter potential is used unchanged in matter with protons and
hyperons, at the density $\tilde n$ of Eq.~\eqref{eq:eint}. It comes from a different interaction than the neutron potential of
Eq.~\eqref{eq:W}.

The factor $\xi_Y$ states how strongly a hyperon contributes to the density that
generates the mean field. No data fix it at the densities needed here, and we treat it as an uncertainty,
$0\le\xi_Y\le1$. For $\xi_Y=1$ a hyperon counts as a nucleon. A further choice is the
fraction of light quarks in $Y$, $2/3$ for $\Lambda$ and $\Sigma$ and $1/3$ for $\Xi$. The
meaning of $\xi_Y$ follows from the expansion at fixed nucleon density $n_N$, here for
one hyperon species in neutron matter:
\begin{equation}
\begin{aligned}
  \varepsilon_{\rm int} ={}& W(n_N) + n_Y U_Y(n_N)\\
   &+ n_Y^2\left[\xi_Y U_Y'(n_N) - \tfrac12\xi_Y^2 W''(n_N)\right] + O(n_Y^3) .
\end{aligned}
  \label{eq:yy}
\end{equation}
The last term is a hyperon--hyperon interaction implied by Eq.~\eqref{eq:eint}. It vanishes
for $\xi_Y=0$. Hyperons then interact only with nucleons, and pure hyperon matter is a free
gas. In model~A (Sec.~\ref{sec:nucl}) with Eq.~\eqref{eq:uy}, $c_\Lambda=0$ and
$\xi_\Lambda=2/3$, take $n_\Lambda=\nz$. This term then gives a $\Lambda$ a potential of $+47$,
$+39$ and $+11\MeV$ at $n_N=2$, $3$ and $4\,\nz$. For a constant $U_Y$ the potential is
$-\xi_Y^2n_YW''$. It is attractive where $W$ is convex, between about $0.6\,\nz$ and $3.8$, $3.1$
and $2.8\,\nz$ in models A, B and C. Double-$\Lambda$ hypernuclei indicate a weak attraction near
$\nz$ \cite{Ahn:2013poa} and give no information at the densities needed here.
Equation~\eqref{eq:eint} is therefore a specified extension of the input potentials and not a
many-body calculation of hyperonic matter. Another functional with the same potentials for
a single hyperon would give a different composition once hyperons are abundant.

\subsection{Nucleonic models}
\label{sec:nucl}

We scanned $\Ls=400$, $350$ and $300\MeV$, $\sigma=2$, $2.5$ and $3$, and $u_c=2.5$,
$3$ and $4$, which gives 27 parameter sets. Table~\ref{tab:models} lists the three
used below. Model~A has the value of $\Ls$ used by FKM. Its quark states saturate at
$2.55\,\nz$, and a maximum mass of $2.1\,\Msun$ is reached only with a neutron
potential of $+200\MeV$ at the centre. In models~B and C the quark states saturate
earlier and the shell carries more of the pressure. The neutron potential at the
centre is $+76$ and $+42\MeV$; in model~B it passes through a maximum of $+106\MeV$
at $3.1\,\nz$.

The chemical potential of Eq.~\eqref{eq:mukin} has an infinite derivative at $\nq$ from
above, because $d\kbu/d\ksh$ diverges as $\kbu\to0$. The squared sound speed therefore exceeds 1 just above
$\nq$, within $0.0025$, $0.0010$ and $0.0003\,\nz$ in models A, B and C. It follows from
the sharp transition of the model, which is therefore not causal in this interval. Outside $1\%$ above $\nq$ the EoS of neutron matter is causal up to
the central density for all 27 sets, with a largest squared sound speed of $0.94$.
The incompressibility of the three models is $389$--$424\MeV$, well above the empirical
range of about $220$--$260\MeV$. The symmetry energy and its slope are $30$--$31\MeV$ and
$67$--$76\MeV$.

A smaller exponent lowers the incompressibility, with $\sigma=4/3$ giving $240$--$290\MeV$
for $u_c\ge3$. At the same neutron potential the maximum mass is then lower. Near
$\UN(\nc)=+200\MeV$ it is $2.01$ against $2.12\,\Msun$ for $\Ls=400\MeV$, and $2.19$ against
$2.32\,\Msun$ for $\Ls=350\MeV$. A small neutron potential with two solar masses is still
possible for $\Ls=300\MeV$. With $\sigma=4/3$ and $u_c=3$, the set has $\Mmax=2.02\,\Msun$,
$\UN(\nc)=+29\MeV$ and an incompressibility of $288\MeV$. For $\Ls=350\MeV$ and
$\sigma=4/3$, $u_c=4$, the values are $1.93\,\Msun$, $+66\MeV$ and $259\MeV$. Section~\ref{sec:req}
applies the condition for the $\Lambda$ to these softer sets.

\begin{table}[t]
\centering
\begin{tabular}{lccc}
\toprule
 & A & B & C \\
\midrule
$\Ls$ (MeV)                         & 400   & 350   & 300 \\
$\sigma$, $u_c$                     & 2, 4  & 2, 3  & 2, 2.5 \\
$K_0$ (MeV)                         & 389   & 405   & 424 \\
$J$, $L$ (MeV)                      & 31, 76 & 30, 72 & 30, 67 \\
$\nq/\nz$                           & 2.55  & 1.71  & 1.08 \\
$\Mmax/\Msun$, neutron matter       & 2.12  & 1.98  & 2.08 \\
$\nc/\nz$                           & 4.74  & 4.53  & 4.22 \\
$\UN(2\nz)$ (MeV)                   & $+86$ & $+62$ & $+42$ \\
$\UN(\nc)$ (MeV)                    & $+200$ & $+76$ & $+42$ \\
$D(\nc)$ (MeV)                      & $-36$ & $-9$  & $+17$ \\
$\UL$ required at $\nc$ (MeV)       & $+164$ & $+67$ & $+58$ \\
same, neutron Fermi gas (MeV)       & $+177$ & $+48$ & $+7$ \\
\bottomrule
\end{tabular}
\caption{Nucleonic models. $K_0$, $J$ and $L$ are the incompressibility, the
symmetry energy and its slope at $\nz$. The other rows refer to pure neutron matter:
$\nq$ is the density at which the $d$-quark states saturate, $\nc$ the central density of
the maximum-mass star and $\UN$ the neutron potential. $D$ is defined in
Eq.~\eqref{eq:cond}. The last two rows give the value of $\UL(\nc)$ above which no $\Lambda$
appears continuously at the centre, in the model and for a neutron Fermi gas with the same
potentials. Each entry is rounded separately.}
\label{tab:models}
\end{table}

\subsection{Methods of solution}
\label{sec:methods}

\emph{FKM occupation probabilities.} With the piecewise-constant functions of
Sec.~\ref{sec:idylliq} the EoS follows in closed form. This is the method of
Ref.~\cite{Fujimoto:2024doc}, extended here by the continuation $k_Y=\kbu$. It assumes that
hyperons occupy only momenta below $\kbu$ and appear only above $\nq$.

\emph{Global minimisation for the $\Lambda$ alone.} For neutrons and $\Lambda$ hyperons at
zero temperature, the kinetic energy density at fixed densities $n_n$, $n_\Lambda$ is a
linear program in the occupation probabilities. We solve it on a grid of $n$ and of the
$\Lambda$ fraction $x_\Lambda=n_\Lambda/n$ from 0 to 1, for each $\Ls$. Adding the interaction
energy and taking the minimum over $x_\Lambda$ needs no iteration and has no dependence on a
starting point.

\emph{Iterative minimisation for the octet.} With protons, leptons and six hyperons we
minimise $\varepsilon$ at fixed baryon density and zero charge over all occupation probabilities
$0\le f_X(k)\le1$ that satisfy Eq.~\eqref{eq:pauli}. The map~\eqref{eq:fq}, the
kernel~\eqref{eq:kernel} and $\Ls$ are kept fixed. A finite-temperature entropy term is used
as a numerical regularisation, and the result at zero temperature is approximated by
calculations at $T=1\MeV$. The occupations are
\begin{equation}
\begin{aligned}
  f_X(k) &= \left[1+e^{\,\omega_X(k)/T}\right]^{-1},\\
  \omega_X(k) &= E_X(k)+U_X+\Phi_X(k)-\muB+Q_X\mue ,
\end{aligned}
  \label{eq:fx}
\end{equation}
with $E_X=\sqrt{k^2+M_X^2}$, the charge $Q_X$ and the electron chemical potential
$\mue$. The function $\Phi_X(k)$ contains the multipliers of the
constraints~\eqref{eq:pauli} and is given in Eq.~\eqref{eq:phi}.
Appendix~\ref{sec:app} gives the algorithms, the tests and the numerical resolution, which
is $0.01\,\Msun$ for the reduction of the maximum mass in the iteration and $0.005\,\Msun$ in the
global minimisation. Where the pressure decreases with density a
Maxwell construction is applied, with local charge neutrality.

\section{Threshold conditions}
\label{sec:thr}

\subsection{$\Lambda$ in neutron matter}
\label{sec:thrL}

Above $\nq$ the $d$-quark constraint is active on the quark momenta below $\kbu/\Nc$. It
adds an energy $\Phi_n(k)$ to a neutron of momentum $k$. A $\Lambda$ has half as many $d$
quarks and feels $\Phi_n/2$. At fixed baryon density the cost of a $\Lambda$ is
\begin{equation}
  \omega_\Lambda(k) = E_\Lambda(k)+\UL-\mukin-\UN+\tfrac12\Phi_n(k) .
  \label{eq:omL}
\end{equation}
The term $\Phi_n/2$ is the statistical repulsion of FKM acting on the $\Lambda$.
Here $\UL$ and $\UN$ are the derivatives of $\varepsilon_{\rm int}$ with respect to the
$\Lambda$ and neutron densities at vanishing $\Lambda$ density. They are those of neutron
matter at the same density and are taken independent of momentum. Appendix~\ref{sec:appmath}
shows that
\begin{equation}
\begin{aligned}
  \Phi_n(k) &= \mukin-E_N(k), \qquad k\le\kbu,\\[2pt]
  \Phi_n(k) &= \Phi_n(\kbu)\,\frac{\kbu}{k}\,e^{-(k-\kbu)/(\Nc\Ls)}, \quad k\ge\kbu .
\end{aligned}
  \label{eq:phin}
\end{equation}
The first line expresses that neutrons with $k<\kbu$ are in equilibrium. The second, with
$\Phi_n(\kbu)$ from the first, holds because the kernel~\eqref{eq:kernel} is the Green's function of a Yukawa operator. The
kernel carries the effect of the quark constraint to baryon momenta above $\kbu$, and it
falls off there.

A $\Lambda$ does not appear continuously at density $n$ if $\omega_\Lambda(k)>0$ for all $k$:
\begin{equation}
\begin{aligned}
  &\UL(n)-\UN(n) > D(n),\\
  &D(n) = \begin{cases}
     \max_k\left[\mukin-E_\Lambda(k)-\tfrac12\Phi_n(k)\right], & n>\nq,\\[2pt]
     E_F(n)-M_\Lambda, & n<\nq .
  \end{cases}
\end{aligned}
  \label{eq:cond}
\end{equation}
The second line is the threshold $\mu_n=M_\Lambda+\UL$ of a neutron gas. We call it the
usual threshold. Apart from the baryon masses, $D$ depends only on $\Ls$.

For $k<\kbu$ the bracket decreases with $k$. The maximum therefore lies either at $k=0$ or
at a momentum $k^*>\kbu$. At $k=0$ the condition reads
\begin{equation}
  \mukin(n) = (2M_\Lambda-M_N) + 2\,\UL(n) - 2\,\UN(n),
  \label{eq:thr}
\end{equation}
with $D_0=\tfrac12(\mukin-2M_\Lambda+M_N)$. For vanishing potentials this is
Eq.~\eqref{eq:fkm}. In terms of $\mukin$ both potentials enter with a coefficient 2. With
$\mu_n=\mukin+\UN$ the same condition reads $\mu_n=(2M_\Lambda-M_N)+2\UL-\UN$, so in terms of
$\mu_n$ the neutron potential has a coefficient 1. Equation~\eqref{eq:thr} holds
at densities where $D_0$ is the larger value. These are $n>3.40$, $2.29$ and $1.45\,\nz$ for
$\Ls=400$, $350$ and $300\MeV$. This includes the threshold of FKM, at $5.74\,\nz$ for
$\UL=0$, which is therefore an exact result for neutron matter with $\Ls=400\MeV$.

Between $\nq$ and those densities a $\Lambda$ enters at $k^*$, just above the saturated
states. $D$ is continuous at $\nq$, because $\kbu$ and $k^*$ vanish there and $\Phi_n$ at fixed
$k>0$ vanishes with $\kbu$. For
$\Ls=400\MeV$ it is $-73\MeV$ at $\nq$, falls to $-93\MeV$ at $2.81\,\nz$ and rises to
$-36\MeV$ at the centre of model~A. The momentum $k^*$ rises steeply above $\nq$ and is $192$ and
$233\MeV$ at $2.75$ and $3\,\nz$. The form $D_0$ alone would give $-124\MeV$ just above $\nq$
and would have a step there. A hyperon at rest replaces half a neutron at rest. Just above
$\nq$ this is more costly than a hyperon outside the saturated states.

Equation~\eqref{eq:cond} was tested against the linear program of
Sec.~\ref{sec:methods}. The dual variables of the program give $D$ and $k^*$ directly. The
two agree within $1.2\MeV$ and the grid spacing for all three values of $\Ls$. The kinetic
energy density of the neutron occupation of FKM agrees with the minimum of the program to
$0.003\,\mathrm{MeV}\,\fm^{-3}$. Appendix~\ref{sec:appmath} proves that the neutron occupation of
FKM is the minimiser in pure neutron matter at every density above $\nq$.

Because $D$ falls above $\nq$, a potential that crosses the required value below $\nq$ can
cross back above it. For Eq.~\eqref{eq:uy} with $c_\Lambda=0$ in model~A, the condition is
violated from $2.36$ to $2.58\,\nz$, holds up to $2.77\,\nz$ and is violated above. At $T=0$ and
for $\xi_\Lambda\ge2/3$ the $\Lambda$ fraction is $1\%$ at $2.5\,\nz$. It is zero at $2.625$ and
$2.75\,\nz$. It is $3.5$--$4\%$ at $2.875\,\nz$ and $8.5$--$9.5\%$ at $3\,\nz$.

Condition~\eqref{eq:cond} concerns the continuous appearance of the $\Lambda$ from
zero density. It does not exclude a first-order transition to a state with a finite
$\Lambda$ density. The global minimisation does not rely on it, and Sec.~\ref{sec:loss}
gives the results.

For constant potentials and $\UN=0$, the case of FKM with a shifted hyperon mass,
the threshold density moves by $0.28\,\nz$ per $10\MeV$ of $\UL$, about twice the
shift of the usual threshold on the same curve $\muB(n)$. The threshold is
$5.74\,\nz$ for $\UL=0$ and $4.96\,\nz$ for $\UL=-28\MeV$.

\subsection{General condition and charged baryons}
\label{sec:thrgen}

The minimisation of Eq.~\eqref{eq:eps} under the constraints~\eqref{eq:pauli} gives
a condition for every species (Appendix~\ref{sec:appmath}). A species $X$ does not
appear if
\begin{equation}
  E_X(k)+U_X-\mu_X+\Nc^3\left[B_u^X\Phat_u(k)+B_d^X\Phat_d(k)\right] > 0
  \label{eq:gen}
\end{equation}
for all $k$, with $\mu_X=\mu_n-Q_X\mue$. The functions $\Phat_u$ and $\Phat_d$ are
the multipliers of the $u$- and $d$-quark constraints, folded with the kernel. In
neutron matter above $\nq$ only the $d$-quark constraint is active and
$\Nc^3B_d^X\Phat_d=\tfrac32B_d^X\Phi_n$. The coefficient of $\Phi_n$ is 1 for the $\Sigma^-$,
$1/2$ for $\Lambda$, $\Sigma^0$ and $\Xi^-$, and 0 for $\Sigma^+$ and $\Xi^0$. A $\Sigma^-$
has the same number of $d$ quarks as a neutron. Its cost is
$E_\Sigma(k)-E_N(k)+\US-\UN-\mue+\omega_n(k)$. With $\omega_n<0$ in the shell, its minimum lies inside
the shell, at a high momentum.

With protons the $u$-quark constraint is active as well, and no closed form is available.
For $\Ls=400\MeV$ and without interaction energy ($W=0$), the linear program
gives the multipliers in neutron--proton--lepton matter in $\beta$ equilibrium, and
Eq.~\eqref{eq:gen} gives the thresholds of Table~\ref{tab:charged}. The form of the occupation
probabilities is not assumed. The $\Xi^0$ and $\Sigma^+$ contain no $d$ quark and enter near
$k=0$. The $\Lambda$ and $\Xi^-$ enter at a finite momentum, and the $\Sigma^-$ and $\Sigma^0$
at a high momentum in the shell. For the $\Lambda$ with protons the threshold moves from $4.96$ to
$5.6\,\nz$ at $\UL=-28\MeV$. FKM estimate the $\Xi^0$ threshold at $5$--$6\,\nz$, close to the values of the
table. The $\Sigma^0$ and $\Sigma^-$ thresholds lie highest. With
these potentials the $\Lambda$ is first, followed closely by the $\Xi^0$.

\begin{table}[t]
\centering
\begin{tabular}{lcccc}
\toprule
 & \multicolumn{2}{c}{vacuum masses} & \multicolumn{2}{c}{$U_Y=(-28,+30,-14)\MeV$} \\
 & $n/\nz$ & $k$ (MeV) & $n/\nz$ & $k$ (MeV) \\
\midrule
$\Lambda$   & 6.7 & 370 & 5.6 & 300 \\
$\Xi^0$     & 6.1 & 0   & 5.9 & 0 \\
$\Xi^-$     & 7.3 & 430 & 7.0 & 420 \\
$\Sigma^+$  & 6.8 & 0   & 7.5 & 0 \\
$\Sigma^0$  & 9.9 & 540 & 11.5 & 600 \\
$\Sigma^-$  & 8.4 & 760 & 9.8 & 830 \\
\bottomrule
\end{tabular}
\caption{Threshold densities and the hyperon momentum at threshold, from the multipliers of the
linear program, for $\Ls=400\MeV$ without interaction energy, in neutron--proton--lepton matter
in $\beta$ equilibrium. The potentials in the last two columns are constant and refer to
$\Lambda$, $\Sigma$ and $\Xi$. Each threshold is computed for matter without other hyperons.
The density resolution is $0.1\,\nz$.}
\label{tab:charged}
\end{table}

\section{Neutron matter with $\Lambda$ hyperons}
\label{sec:neutral}

\subsection{The required $\Lambda$ potential}
\label{sec:req}

\begin{figure}[t]
\includegraphics[width=\columnwidth]{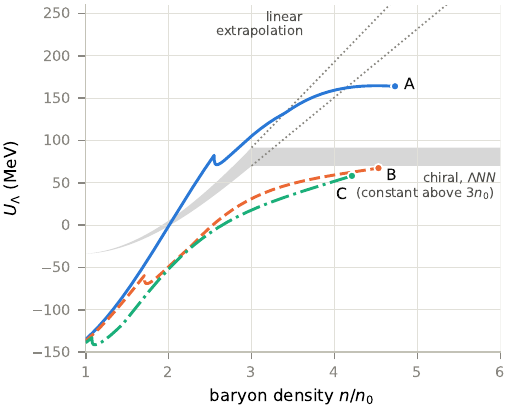}
\caption{$\UN(n)+D(n)$ of Eq.~\eqref{eq:cond} for models~A (solid), B (dashed) and C
(dash-dotted). For a $\Lambda$ potential above the curve no $\Lambda$ appears continuously at
density $n$. Each curve ends at the central density of the maximum-mass star. The
kink is at the quark saturation density $\nq$. Grey band: $\Lambda$ potential in
neutron matter from chiral effective field theory with $\Lambda NN$ forces
\cite{Gerstung:2020ktv}, kept constant above $3\,\nz$. Dotted: linear extrapolation
of its two edges.}
\label{fig:req}
\end{figure}

Figure~\ref{fig:req} shows $\UN(n)+D(n)$ for the three models. No $\Lambda$ appears
continuously in the star if the actual potential lies above the curve up to the central density.

For model~A the required potential is $-1\MeV$ at $2\,\nz$, $+76\MeV$ at $2.5\,\nz$,
$+105\MeV$ at $3\,\nz$ and $+164\MeV$ at the centre. Just above $\nq$ it dips to $+71\MeV$
at $2.58\,\nz$. The two edges of the chiral-EFT band cross the curve at $2.0$ and $2.1\,\nz$. The
$\Lambda$ therefore appears by the usual threshold, before the quark states saturate. The
constant band stays below the curve up to the centre. With the linear extrapolation it
returns above the curve at $4.1$ and $3.4\,\nz$.

Equation~\eqref{eq:uy} with $\UL(\nz)=-28\MeV$ and $c_\Lambda=0$ gives $+98\MeV$ at
$3\,\nz$ in the neutron matter of model~A, above the chiral-EFT band and below the required
value. The condition holds up to the centre for $c_\Lambda\ge3.3\MeV$. Matching
Eq.~\eqref{eq:uy} to the chiral-EFT band at $3\,\nz$ would require $c_\Lambda$ between $-4.7$ and
$-1.1\MeV$.

For models~B and C the required potential at the centre is $+67$ and $+58\MeV$. The
chiral-EFT band lies above both curves at all densities, for the constant and for the
linear extrapolation. For the constant extrapolation the lower edge clears the curve by only
$3$ and $12\MeV$ at the centre. Equation~\eqref{eq:uy} with $c_\Lambda=0$ is less repulsive
there than the chiral-EFT band and falls below the curves by $12$ and $21\MeV$. It satisfies the
condition for $c_\Lambda\ge0.7$ and $1.5\MeV$.

Table~\ref{tab:models} compares with a neutron Fermi gas with the same potentials.
In model~A the saturated quark states lower the required $\UL(\nc)$ by $13\MeV$. In
models~B and C they raise it, because $\mukin$ rises faster than the Fermi energy.
The condition holds in models~B and C because the neutron potential is small.
There the condition~\eqref{eq:cond} requires a larger $\Lambda$ potential than the
usual one.

The same analysis was made for all 27 parameter sets, in neutron matter. With the chiral-EFT band
held constant above $3\,\nz$ the condition holds for both edges when $\UN(\nc)$ is below about
$+80\MeV$. The six sets that satisfy this have maximum masses between $1.55$ and $2.08\,\Msun$.
A seventh, with $\Mmax=1.98\,\Msun$ and $\UN(\nc)=+82\MeV$, satisfies it for the upper edge only.
Every set with $\Mmax>2.1\,\Msun$ has $\UN(\nc)>+89\MeV$ and violates the condition for at least
one edge. With the linear extrapolation three further sets satisfy it for both edges. Their
maximum masses are $1.98$, $2.24$ and $2.24\,\Msun$, and the margin in the last two is
$2$--$18\MeV$. The value of $+80\MeV$ summarises this scan and is not a general bound. It does
not replace Eq.~\eqref{eq:cond}, which must hold at every density; $\UN(n)$ need not be
monotonic, as model~B shows.

The softer sets of Sec.~\ref{sec:nucl}, with $\sigma<2$, were tested with the condition alone. No
minimisation was made for them. With the
constant extrapolation three of them satisfy the condition for both edges. They have
$\Ls=300\MeV$ and $u_c=3$. Their maximum masses are $2.02$--$2.11\,\Msun$, and their
incompressibilities are $288$--$346\MeV$, above the empirical range. In every set that satisfies
the condition for both constant edges, $\UN(\nc)$ is below $+80\MeV$. With the linear
extrapolation more sets satisfy it, up to $2.25\,\Msun$. Three of these have an incompressibility
of $243$ or $259\MeV$. They reach $1.93$, $2.01$ and $2.18\,\Msun$. The second has $\Ls=400\MeV$,
$\sigma=4/3$ and $u_c=6$, with $\UN(\nc)=+198\MeV$. Its lower edge clears the condition by $7\MeV$.
With the threshold of a neutron Fermi gas in place of $D$ it would fail by $16\MeV$. With the
constant extrapolation the $\Lambda$ appears in this set at $3.3$--$3.5\,\nz$. The tidal
deformabilities of the softer sets were not computed.

\subsection{Reduction of the maximum mass}
\label{sec:loss}

Table~\ref{tab:loss} gives the reduction of the maximum mass, $\Delta\Mmax$, relative to the
star without hyperons computed in the same way. The FKM occupation probabilities, computed with
$\xi_\Lambda=0$, give at most $0.01\,\Msun$ in model~A. For the chiral potentials the $\Lambda$
threshold lies below $\nq$ there, outside the range of validity stated by FKM
(Sec.~\ref{sec:idylliq}). The other columns come from the global
minimisation over all occupation probabilities and all $\Lambda$ fractions.

\begin{table*}[t]
\centering
\begin{ruledtabular}
\begin{tabular}{lcccccc}
 & FKM occ. & \multicolumn{5}{c}{global minimisation} \\
$\Lambda$ potential & & $\xi_\Lambda=0$ & $0.2$ & $0.4$ & $2/3$ & $1$ \\
\colrule
\multicolumn{7}{l}{Model A}\\
Eq.~\eqref{eq:uy}, $c_\Lambda=0$          & $<0.005$ & 0.34 & 0.19 & 0.09 & 0.06 & 0.03 \\
Eq.~\eqref{eq:uy}, $c_\Lambda=2.5\MeV$    & $<0.005$ & 0.33 & 0.17 & 0.04 & $<0.005$ & $<0.005$ \\
Eq.~\eqref{eq:uy}, $c_\Lambda=5\MeV$      & $<0.005$ & 0.33 & 0.15 & $<0.005$ & $<0.005$ & $<0.005$ \\
Eq.~\eqref{eq:uy}, $c_\Lambda=10\MeV$     & $<0.005$ & 0.32 & 0.13 & $<0.005$ & $<0.005$ & $<0.005$ \\
chiral, constant, lower edge              & 0.01 & 0.46 & 0.36 & 0.28 & 0.26 & 0.16 \\
chiral, constant, upper edge              & 0.01 & 0.44 & 0.32 & 0.20 & 0.15 & 0.11 \\
chiral, linear, lower edge                & $<0.005$ & 0.46 & 0.36 & 0.28 & 0.07 & $<0.005$ \\
chiral, linear, upper edge                & $<0.005$ & 0.44 & 0.32 & 0.20 & 0.01 & $<0.005$ \\
\colrule
\multicolumn{7}{l}{Model B}\\
Eq.~\eqref{eq:uy}, $c_\Lambda=0$ to $10\MeV$ & n.c. & 0.09--0.10 & 0.01 & $<0.005$ & $<0.005$ & $<0.005$ \\
chiral, constant or linear, lower edge    & n.c. & 0.14 & 0.06 & $<0.005$ & $<0.005$ & $<0.005$ \\
chiral, constant or linear, upper edge    & n.c. & 0.13 & 0.04 & $<0.005$ & $<0.005$ & $<0.005$ \\
\colrule
\multicolumn{7}{l}{Model C}\\
Eq.~\eqref{eq:uy}, $c_\Lambda=0$ to $10\MeV$ & n.c. & 0.04 & 0.01 & $<0.005$ & $<0.005$ & $<0.005$ \\
chiral, constant or linear                & n.c. & 0.06 & 0.02 & $<0.005$ & $<0.005$ & $<0.005$ \\
\end{tabular}
\end{ruledtabular}
\caption{Reduction of the maximum mass in $\Msun$ for neutron matter with $\Lambda$ hyperons, without
a crust. ``FKM occ.''\ uses the FKM occupation probabilities, which admit hyperons only above the
quark saturation density and below the momentum $\kbu$, computed with $\xi_\Lambda=0$ (n.c.:
not computed). For the chiral potentials in model~A this is outside the range stated by FKM. The other columns are the global minimisation at $T=0$ of
Sec.~\ref{sec:methods}, with the maximum masses without hyperons of $2.12$, $1.98$ and
$2.08\,\Msun$. ``Chiral'' is the band of Ref.~\cite{Gerstung:2020ktv} up to $3\,\nz$ with the
constant or linear extrapolation above.}
\label{tab:loss}
\end{table*}

The large values at small $\xi_\Lambda$ follow from Eq.~\eqref{eq:yy}. When a hyperon does not
contribute to $\tilde n$, converting a neutron lowers $\tilde n$ and with it both $W$ and $U_Y$.
A phase with most baryons in $\Lambda$ hyperons is then favoured even for a large
$c_\Lambda$. In model~A with $c_\Lambda=10\MeV$ the condition~\eqref{eq:cond} holds up to the
centre. Yet at $\xi_\Lambda=0$ a first-order transition sets in at $2.6\,\nz$, with a density jump to
$4.2\,\nz$, and the reduction is $0.32\,\Msun$. The condition is therefore not sufficient for the absence
of hyperons. In the same functional the FKM occupation probabilities cannot reach this phase, since
they limit the hyperons to occupation $1/9$ below $\kbu$.

At $\xi_\Lambda\ge0.4$ the term in $\xi_\Lambda U_Y'$ of Eq.~\eqref{eq:yy} is larger. For
Eq.~\eqref{eq:uy} the reduction falls to $0.03$--$0.09\,\Msun$ at $c_\Lambda=0$ and vanishes for
$c_\Lambda\ge5\MeV$. This is the behaviour expected from the condition. Two effects are absent
from the FKM occupation probabilities and matter for all $\xi_\Lambda$. First, the $\Lambda$ appears
below $\nq$ when the usual threshold is met there. Second, above $\nq$ the $\Lambda$ occupies
momenta above $\kbu$, where the quark states are not saturated, with occupation up to 1.

With the chiral potential held constant above $3\,\nz$, $\UN-\UL$ exceeds $100\MeV$ at
$4.74\,\nz$, the central density without hyperons. For $\xi_\Lambda\ge2/3$ the reduction
remains $0.11$--$0.26\,\Msun$. The $\Lambda$ fraction at the centre is $50$--$58\%$ for
$\xi_\Lambda=2/3$, at $4.2$--$4.3\,\nz$. It is $36$--$45\%$ for $\xi_\Lambda=1$, at $5.3$--$6.8\,\nz$.
The first density jump is from $2.7$--$3.2$ to $3.4$--$4.0\,\nz$.
It lies near the kink of the constant extrapolation at $3\,\nz$. Replacing the kink by a slope
that falls linearly to zero over $\pm0.25\,\nz$ changes the reduction by less than $0.005\,\Msun$.
Over $\pm0.5\,\nz$ the reduction grows by up to $0.02\,\Msun$ and the jump near $3\,\nz$ disappears at
both values of $\xi_\Lambda$. The size of the reduction therefore does not depend on the kink. The
first-order character of this transition does.

With the linear extrapolation and $\xi_\Lambda\ge2/3$ the reduction is at most $0.07\,\Msun$. The repulsion
above $3\,\nz$ then does what it does in a hadronic model. The star is still not free of hyperons. For $\xi_\Lambda=1$ the $\Lambda$ fraction
reaches $18$--$38\%$ near $3\,\nz$ and vanishes above $3.5$--$4\,\nz$. For $\xi_\Lambda=2/3$ it
grows up to $63$--$83\%$, and at fixed density the minimum returns to neutron matter at $4.25$ and $5.25\,\nz$.

Without interaction energy the same comparison can be made for the model of FKM
itself. With a constant $\UL=-28\MeV$ the FKM occupation probabilities give
$\Delta\Mmax=0.04\,\Msun$. The iterative minimisation gives $0.10\,\Msun$.

For $\xi_\Lambda\ge0.4$ no $\Lambda$ is present in the stars of models~B and C for the chiral
potentials. For Eq.~\eqref{eq:uy} with $c_\Lambda=0$ the $\Lambda$ fraction at the centre is
$1$--$3\%$. In both cases the reduction is below the resolution, and the minimisation finds no
$\Lambda$-rich state inside the star. At $\xi_\Lambda=0.2$ the reduction is $0.01$--$0.06\,\Msun$. At $\xi_\Lambda=0$ it is up to $0.14$ and
$0.06\,\Msun$. The cause is a transition at $2.8$ and $3.0\,\nz$ to a phase with more than half of
the baryons in $\Lambda$ hyperons.

\section{Baryon octet in $\beta$ equilibrium}
\label{sec:octet}

The iterative minimisation is now applied to neutrons, protons, the $\Lambda$, the
three $\Sigma$, the two $\Xi$, electrons and muons, with the constraints for $u$,
$d$ and $s$ quarks. All hyperon potentials follow Eq.~\eqref{eq:uy} with a common
coefficient $c_Y$, or the $\Lambda$ potential is replaced by the chiral-EFT band while
$c_\Sigma=c_\Xi=0$. The $\Sigma$ and $\Xi$ potentials above $\nz$ are therefore
always those of Eq.~\eqref{eq:uy}. We use three choices: $\xi_Y=0.4$ for all hyperons, the
light-quark fraction, and $\xi_Y=1$. Table~\ref{tab:octet} gives the results.

At smaller $\xi_Y$ the iteration has many solutions, and it does not reliably reach the
lowest one. No global minimisation over the octet was made. A restricted check with neutrons and
the $\Xi^0$ alone finds no $\Xi^0$-rich phase at any $\xi$. At $\xi_Y=0$ the iteration gives
a reduction of $0.11$--$0.44\,\Msun$ in model~A. In model~C, and in model~B except for $c_Y=0$, it
misses the $\Lambda$-rich phase of Sec.~\ref{sec:loss}. The reductions for $\xi_Y=0$ and $0.2$ in Table~\ref{tab:loss} are
therefore the only values for small $\xi$ that we regard as established.

\begin{table*}[t]
\centering
\begin{ruledtabular}
\begin{tabular}{lccc}
 & Model A & Model B & Model C \\
\colrule
$\Mmax$ without hyperons ($\Msun$)            & 2.11 & 1.97 & 2.05 \\
proton fraction, $\mue$ at the centre, without hyperons & 9\%, $213\MeV$ & 6\%, $187\MeV$ & 6\%, $178\MeV$ \\
\colrule
$\Delta\Mmax$, $c_Y=0$                        & 0.07--0.08 & $<0.01$ & $<0.01$ \\
$\Delta\Mmax$, $c_Y=2.5\MeV$                  & 0.03--0.04 & $<0.01$ & $<0.01$ \\
$\Delta\Mmax$, $c_Y=5\MeV$                    & $\le0.01$ & $<0.01$ & $<0.01$ \\
$\Delta\Mmax$, $c_Y=10\MeV$                   & $<0.01$ & $<0.01$ & $<0.01$ \\
$\Delta\Mmax$, chiral $\UL$, constant         & 0.15--0.30 & $<0.01$ & $<0.01$ \\
$\Delta\Mmax$, chiral $\UL$, linear           & 0.07--0.24 & $<0.01$ & $<0.01$ \\
\colrule
first hyperon, $c_Y=0$                        & $\Xi^-$ at $2.75\,\nz$ & $\Xi^0$ at $4.75\,\nz$ & $\Xi^0$ at $4.0\,\nz$ \\
hyperons at the centre, $c_Y=0$               & 20--21\% $\Xi^-$ & none & 2.5--3\% $\Xi^0$ \\
\end{tabular}
\end{ruledtabular}
\caption{Baryon octet with electrons and muons in $\beta$ equilibrium, iterative
minimisation at $T=1\MeV$, without a crust. $\Delta\Mmax$ is the reduction of the maximum mass in $\Msun$
relative to neutron--proton--lepton matter. Ranges cover $\xi_Y=0.4$, the light-quark fraction
and $\xi_Y=1$ and, for the chiral potentials, the two edges of the band. Solutions from a cold
and a hot start enter the envelope of $P(\muB)$. Values below the numerical resolution are given
as $<0.01$. The first hyperon is given at the resolution of the density grid, $0.25\,\nz$, for
a fraction above $10^{-3}$; in model~B it lies beyond the central density $4.5\,\nz$.}
\label{tab:octet}
\end{table*}

\begin{figure}[!htb]
\includegraphics[width=\columnwidth]{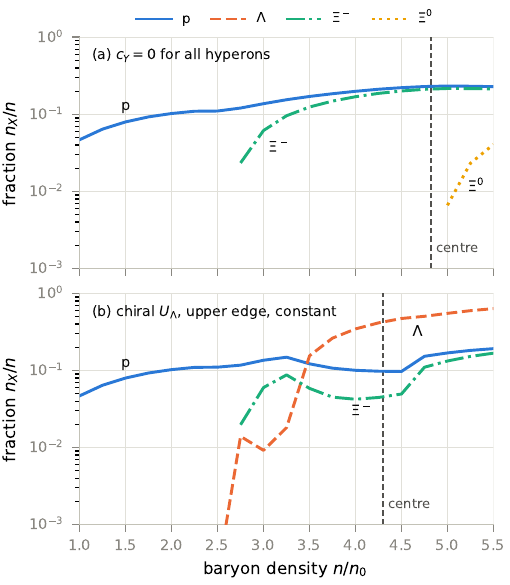}
\caption{Particle fractions in model~A with the full octet in $\beta$ equilibrium,
for $\xi_Y$ equal to the light-quark fraction. (a)~All hyperon potentials from
Eq.~\eqref{eq:uy} with $c_Y=0$. (b)~$\Lambda$ potential from the upper edge of the
chiral-EFT band, constant above $3\,\nz$. The vertical line marks the central density of
the maximum-mass star. The points are solutions at fixed baryon density on a grid
of $0.25\,\nz$; in (b) the Maxwell construction replaces the interval $3.3$--$3.6\,\nz$
by a density jump.}
\label{fig:comp}
\end{figure}

In model~A with $c_Y=0$ the first hyperon is the $\Xi^-$, between $2.5$ and
$2.75\,\nz$ (Fig.~\ref{fig:comp}). It enters by the usual threshold
$\mu_n+\mue=M_{\Xi^-}+\UX$. At $2.5\,\nz$ the electron chemical potential is
$186\MeV$, which nearly compensates the mass difference of $206\MeV$ between
$\Xi^-$ and $\Lambda$. Equation~\eqref{eq:uy} gives $\UX=+36\MeV$ and
$\UL=+72\MeV$ there, because the $\Xi$ has one light quark and the $\Lambda$ two.
The $\Xi^-$ threshold is then $16\MeV$ lower than that of the $\Lambda$. The
$\Xi^-$ fraction reaches $20$--$21\%$ at the centre. Charge neutrality raises the proton
fraction to $22$--$23\%$, and no $\Lambda$ is present at the centre. The reduction is
$0.07$--$0.08\,\Msun$. It falls to $0.03$--$0.04\,\Msun$ for $c_Y=2.5\MeV$ and is
below the resolution for $c_Y=10\MeV$. This result rests on the assumed $\UX$ above
$\nz$, for which no data exist.

With the constant extrapolation of the chiral $\Lambda$ potential the $\Lambda$
fraction at $4.75\,\nz$ is $38$--$63\%$ and the reduction is $0.15$--$0.30\,\Msun$. For
the lower edge of the band the $\Lambda$ is then the first hyperon, at $2.25\,\nz$. The
dependence on $\xi_Y$ of Sec.~\ref{sec:loss} applies. In $\beta$ equilibrium a further property
of Eq.~\eqref{eq:eint} acts in the same direction. The asymmetry $\delta$ contains the nucleon
densities only. Converting neutrons to hyperons at a fixed proton density lowers $\delta$ and
with it the symmetry energy. No $\Sigma$ appears in any case. The $\Sigma$ potential of
Eq.~\eqref{eq:uy} is $+217\MeV$ at $2.5\,\nz$. Its repulsion at $\nz$ is empirical; its
growth with density is an assumption.

In model~B no hyperon is present in the maximum-mass star for $c_Y\ge0$ or for the
chiral $\Lambda$ potentials. The first hyperon, the $\Xi^0$, would appear at
$4.75\,\nz$, beyond the central density. In model~C the $\Xi^0$ appears at $4.0\,\nz$
for $c_Y=0$ and makes up $2.5$--$3\%$ of the centre. In both models the reduction is below
the resolution for the values of $\xi_Y$ of Table~\ref{tab:octet}.

Without interaction energy, and with constant potentials $(-28,+30,-14)\MeV$, the
$\Lambda$ fraction exceeds $10^{-3}$ at $6\,\nz$ and the $\Xi^0$ fraction at
$6.5\,\nz$, on a density grid of $0.5\,\nz$. No $\Sigma$ and no $\Xi^-$ is present
up to the centre at $8.6\,\nz$. The reduction is $0.10\,\Msun$ on a maximum mass of
$1.44\,\Msun$, against $1.46\,\Msun$ for
neutron matter in closed form (Sec.~\ref{sec:intro}).

\section{Discussion}
\label{sec:disc}

\begin{table*}[t]
\centering
\begin{ruledtabular}
\begin{tabular}{p{0.47\textwidth}p{0.47\textwidth}}
Result & Assumptions \\
\colrule
No $\Lambda$ appears continuously if $\UL-\UN>D(n)$ up to the centre, with $D$ in closed form,
Eqs.~\eqref{eq:phin} and \eqref{eq:cond}. The threshold of FKM, Eq.~\eqref{eq:thr}, is the limit
in which the maximum lies at $k=0$. &
Quark Pauli blocking constraints of FKM; neutron matter; potentials independent of momentum,
evaluated at the same density. A first-order transition to a finite $\Lambda$ density is
not excluded. \\
With the chiral $\Lambda$ potential held constant above $3\,\nz$, the condition holds only for
$\UN(\nc)\lesssim+80\MeV$; such parameter sets reach at most $2.11\,\Msun$. With the linear
extrapolation it holds up to $2.25\,\Msun$, and at $2.01\,\Msun$ for a set with $\Ls=400\MeV$ and an
incompressibility of $243\MeV$. &
Nucleonic functional~\eqref{eq:W}, 27 parameter sets and softer sets with $\sigma<2$; neutron
matter; band of Ref.~\cite{Gerstung:2020ktv}, extrapolated; condition~\eqref{eq:cond} only. \\
The reduction of the maximum mass is not limited to the value obtained with the FKM occupation
probabilities. In model~A with the $\Lambda$ alone it is $0.11$--$0.46\,\Msun$ for the chiral potential held constant, up to $0.46$
and at most $0.07\,\Msun$ for $\xi\ge2/3$ with the linear extrapolation, and $0.03$--$0.34\,\Msun$ for
Eq.~\eqref{eq:uy} with $c_\Lambda=0$. For the octet with the three choices of $\xi_Y$ it is $0.07$--$0.30\,\Msun$
for $c_Y=0$ or chiral $\UL$. &
Global minimisation at $T=0$ for the $\Lambda$, iteration at $T=1\MeV$ for the octet;
interaction energy~\eqref{eq:eint} with $\xi$ between 0 and 1 (octet: $\xi_Y$ of 0.4, 1 or the light-quark fraction); for the
octet, $\US$ and $\UX$ from Eq.~\eqref{eq:uy} with $c_Y=0$; Maxwell construction, exact for the
$\Lambda$ alone. \\
In models~B and C the star contains no $\Lambda$ or a few per cent of hyperons, and the reduction is
below $0.01\,\Msun$, for $\xi\ge0.4$. For $\xi=0$ the reduction with the $\Lambda$ alone is up to $0.14$
and $0.06\,\Msun$. &
As above, with $c_Y\ge0$ or chiral $\UL$; $\US(\nz)=+30\MeV$, $\UX(\nz)=-14\MeV$; no crust. \\
\end{tabular}
\end{ruledtabular}
\caption{Main results and the assumptions on which they rest.}
\label{tab:summary}
\end{table*}

\subsection{Comparison with hadronic models}
\label{sec:hadronic}

In a hadronic model with momentum-independent potentials and free masses the $\Lambda$ does not
appear if $\UL-\UN>E_F-M_\Lambda$, the second line of Eq.~\eqref{eq:cond}. In both descriptions the $\Lambda$ must feel
most of the repulsion that the neutron feels. Quarkyonic matter does not remove this
requirement. It changes how much repulsion the neutron needs. In a hadronic model
the pressure above $2\,\nz$ comes mainly from the interaction, and the neutron
potential at the centre is large. In models~B and C part of the pressure comes from
the baryon shell, $\UN$ at the centre is below $+80\MeV$, and the chiral $\Lambda$
potentials satisfy the condition.

Model~A is close to a hadronic model in this respect. Relativistic
mean-field calculations with the parameter sets GM1 \cite{glendenning1991} and FSU2H
\cite{Tolos:2016hhl} give reductions of $0.31$ and $0.16\,\Msun$ relative to the same models without the
$\Lambda$, with $\Lambda$ potentials of $+98$ and $+96\MeV$ at $3\,\nz$ in neutron matter. These values
were computed with the mean-field code of Ref.~\cite{companionPRD}, for stars in $\beta$ equilibrium
with a crust, and the reference without $\Lambda$ still contains $\Sigma^-$ hyperons. The same two
$\Lambda$ potentials in the neutron matter of model~A, with the global minimisation and without a
crust, give $0.37$--$0.43\,\Msun$ at $\xi_\Lambda=0$, $0.16$--$0.17\,\Msun$ at $\xi_\Lambda=0.4$
and less than $0.005\,\Msun$ at $\xi_\Lambda\ge2/3$. The mean-field reductions lie inside this range.
A comparison at the same $\UL(n)$ therefore does not separate the two
descriptions. The outcome also depends on the neutron potential and on the hyperon--hyperon
interaction.

The hyperon potentials are measured at and below $\nz$. Hypernuclei and $\Sigma$
atoms fix them at $\nz$ \cite{gal2016}, and hyperon production by neutrinos in nuclei
is sensitive to $\UL$ near $0.8\,\nz$ \cite{companionPRD}. Neither constrains the
potentials above $3\,\nz$, which decide the results for model~A. The chiral
calculations reach $3.5\,\nz$, and the two extrapolations used here differ by up to
$0.2\,\Msun$ in the reduction at $\xi_\Lambda=2/3$. The neutron potential above $3\,\nz$ is not measured
either. It separates model~A from models~B and C. These two and $\xi$ are the largest
uncertainties of the present results. The quarkyonic part of the condition, $D(n)$, is known in
closed form once $\Ls$ is chosen.

\subsection{Limitations}
\label{sec:lim}

The kernel~\eqref{eq:kernel} and the constraints~\eqref{eq:pauli} are those of the
model of FKM and are not derived from QCD. The nucleonic functional~\eqref{eq:W} is
schematic and is not one of the functionals in use for neutron-star matter. Neither it nor
Eq.~\eqref{eq:eint} is derived from the quarkyonic model. They add a phenomenological
interaction energy to the kinetic energy of FKM. The
incompressibility is $389$--$424\MeV$. The potentials are independent of momentum, although
hyperons enter at momenta up to the top of the shell. The hyperon potentials have no isovector
part, so the charge states of the $\Sigma$ and of the $\Xi$ feel the same potential. Hyperons interact with each other only
through Eq.~\eqref{eq:eint}. First-order transitions are treated with a Maxwell construction.
For neutrons and $\Lambda$ hyperons alone it is exact, because there is a single conserved charge.
With charged particles a Gibbs construction could change the reductions of the maximum mass
(Appendix~\ref{sec:appmath}).
In model~C the quark states saturate at $1.08\,\nz$ in neutron matter and at $1.09\,\nz$ in
symmetric matter, just above the density of nuclei.

The maximum masses of Tables~\ref{tab:models} and \ref{tab:octet} have no crust. With the SLy
crust and $\beta$-equilibrated matter the models have $R_{1.4}=13.3$--$14.1$~km. Their tidal
deformabilities $\Lambda_{1.4}$ are about $790$, $810$ and $1100$. Their pressures at $2\,\nz$ are
$34$, $37$ and $44\,\mathrm{MeV}\,\fm^{-3}$. The analysis of GW170817 gives
$\Lambda_{1.4}=190^{+390}_{-120}$, component radii of $11.9\pm1.4$~km and about
$22^{+17}_{-11}\,\mathrm{MeV}\,\fm^{-3}$ for that pressure \cite{abbott2018eos}. The tidal
deformabilities of all three models lie above this interval. The pressure of models~A and B lies inside the interval, and that of
model~C above it.
In $\beta$ equilibrium models~B and C reach $1.97$ and $2.05\,\Msun$ without a crust and $1.96$ and
$2.04\,\Msun$ with it, below or at the lower end of the mass of PSR~J0740$+$6620 \cite{fonseca2021}. They are examples within this family of
functionals. They show that hyperons can be absent in a quarkyonic model with a stiff shell.
They do not establish an EoS that is compatible with nuclear data and with the measured
masses.

\subsection{Related work}
\label{sec:related}

The review by Kojo \cite{Kojo:2026bnm} keeps vacuum masses and neglects $n$--$\Lambda$
interactions. Interactions have been added to quarkyonic matter for nucleons only in
Refs.~\cite{Saito:2025yld,Folias:2024upz,Folias:2025zmu}, and isospin asymmetry in
Refs.~\cite{Margueron:2021dtx,Zhao:2020dvu}. References~\cite{Duarte:2020xsp,Duarte:2020kvi} use
an excluded-volume description with nucleons, $\Lambda$ hyperons and quarks; the $\Lambda$ forms a
shell only when its size is smaller than that of the nucleon. Reference~\cite{Sun:2026vxk} treats
quarkyonic matter with the hyperon octet and strange quarks in a relativistic mean-field model,
with quarks and baryons in separate Fermi seas, and finds a softening near twice $\nz$.
Reference~\cite{Hu:2026quarkyonic} uses the full octet with a Skyrme functional and separate
quark seas distinct from the constituent quarks of the baryons. There the quarkyonic mechanism
lowers the hyperon critical density, because the nucleons are displaced to higher momentum, and
raises the maximum mass. Pauli blocking of hyperons by quarks has been studied in two-colour QCD
\cite{Nagatsuka:2025ijc}. Reference~\cite{Gao:2026scv} suppresses hyperons through the
chiral-invariant mass, without quarkyonic matter.

\section{Summary}
\label{sec:summary}

Table~\ref{tab:summary} lists the main results with their assumptions.

The threshold of FKM generalises to Eq.~\eqref{eq:cond} when the baryons feel
single-particle potentials. In neutron matter it is exact where the hyperon enters at rest, and
elsewhere the hyperon enters at a momentum given by Eq.~\eqref{eq:phin}. Reference~\cite{Fujimoto:2024doc}
states that the quark Pauli blocking constraints delay the $\Lambda$ and $\Sigma^0$ to $5$--$6\,\nz$ and that they
limit the softening by these hyperons. Those statements refer to vacuum masses and to the occupation
probabilities of FKM. In the models studied here the delay holds for the chiral potentials when the
baryon shell supplies the
stiffness, and then because the neutron potential is small. The softening is limited in the same
models for $\xi\ge0.4$. In model~A, where the stiffness comes from nucleonic repulsion, the
$\Lambda$ appears near $2\,\nz$ for chiral potentials. A small reduction then
needs the $\Lambda$ potential to keep rising above $3\,\nz$ and $\xi\ge2/3$, as in a hadronic model.
With the octet it also needs additional repulsion for the $\Xi$. With the constant extrapolation,
the sets of the functional~\eqref{eq:W} that
satisfy the condition with a star above $1.9\,\Msun$ have $\Ls\le350\MeV$, which corresponds to
quark saturation at or below $1.7\,\nz$. With the linear extrapolation a softer set with
$\Ls=400\MeV$ and an incompressibility of $243\MeV$ satisfies it at $2.01\,\Msun$. For that set no
minimisation was made and the tidal deformability was not computed. Whether a model of this kind
can reproduce nuclear data and reach the measured masses remains to be shown.

\begin{acknowledgments}
This work was supported by the Science and Technology Facilities Council (STFC)
Lancaster EPP Consolidated Grant 2025--2029: UKRI2846.
\end{acknowledgments}

\section*{Data availability}
The programs and the numerical results underlying this work are available from the author on
reasonable request.

\appendix

\section{Derivations}
\label{sec:appmath}

\subsection{Quark occupation}

With $p=k/\Nc$, the average of the kernel~\eqref{eq:kernel} over the angle between
$\mathbf{q}$ and $\mathbf{k}$ is
\begin{equation}
  \overline{K}(q,p) = \frac{e^{-|q-p|/\Ls}-e^{-(q+p)/\Ls}}{8\pi\Nc^3\Ls\,q\,p},
  \label{eq:kbar}
\end{equation}
and $f_q(q)=\Nc^3\sum_X B_q^X\int 4\pi k^2dk\,\overline{K}(q,k/\Nc)f_X(k)$. For
$q\to0$, $\overline{K}\to K(p)$. A sphere filled up to $k_F$ gives
\begin{equation}
\begin{aligned}
  f_q(0) &= B_q^X\Nc^3\,G(k_F),\\
  G(k) &= 1-e^{-x}(1+x),\qquad x=\frac{k}{\Nc\Ls}.
\end{aligned}
  \label{eq:G}
\end{equation}
In neutron matter $f_d(0)=18\,G(k_F)$, and $G(\kq)=1/18$ defines $\kq$.

Above $\nq$ the neutron occupation is $1/18$ for $k<\kbu$ and 1 for $\kbu<k<\ksh$.
The condition $f_d(0)=1$ reads $G(\kbu)+18[G(\ksh)-G(\kbu)]=1$, or
\begin{equation}
  17\,(1+x_{\rm bu})\,e^{-x_{\rm bu}} = 18\,(1+x_{\rm sh})\,e^{-x_{\rm sh}} ,
  \label{eq:kbu}
\end{equation}
which fixes $\kbu(\ksh)$. The density and the kinetic energy density are
$n=\nu(\ksh)-\tfrac{17}{18}\nu(\kbu)$ and
$\varepsilon_{\rm kin}=e(\ksh)-\tfrac{17}{18}e(\kbu)$, where $\nu(k)=k^3/3\pi^2$ and
$e(k)$ are the density and energy density of a filled sphere. With
$r=d\kbu/d\ksh=\ksh(\Nc\Ls+\kbu)/[\kbu(\Nc\Ls+\ksh)]$ from Eq.~\eqref{eq:kbu},
\begin{equation}
  \mukin = \frac{d\varepsilon_{\rm kin}}{dn}
   = \frac{\ksh^2E_N(\ksh)-\tfrac{17}{18}\,r\,\kbu^2E_N(\kbu)}
          {\ksh^2-\tfrac{17}{18}\,r\,\kbu^2}.
  \label{eq:mukin}
\end{equation}

\subsection{Single-particle potentials}

Let $V_Y=U_Y-\xi_Y\,\partial W/\partial\tilde n$. Equation~\eqref{eq:eint} gives
\begin{equation}
\begin{aligned}
  U_{n,p} &= \frac{\partial W}{\partial n_{n,p}}
     + \sum_Y n_Y\,\frac{\partial V_Y}{\partial n_{n,p}},\\
  \frac{\partial\varepsilon_{\rm int}}{\partial n_Y} &= U_Y(\tilde n,\delta)
     + \xi_Y\sum_{Y'} n_{Y'}\,\frac{\partial V_{Y'}}{\partial\tilde n}.
\end{aligned}
  \label{eq:sp}
\end{equation}
The sums are the rearrangement terms. They vanish when no hyperon is present. The
potentials in the threshold conditions are therefore $\partial W/\partial n_n$ and
the input $U_Y$. Since all potentials are derivatives of one function,
$P=\sum_X\mu_Xn_X+\mue(n_e+n_\mu)-\varepsilon=\muB n-\varepsilon$ in $\beta$
equilibrium at zero charge. Expanding Eq.~\eqref{eq:eint} in $n_Y$ at fixed $n_N$, with
$\tilde n=n_N+\xi_Yn_Y$, gives Eq.~\eqref{eq:yy}: the terms $\xi_Yn_Y W'$ cancel at first order,
and at second order $\tfrac12\xi_Y^2n_Y^2W''+\xi_Yn_Y^2U_Y'-\xi_Y^2n_Y^2W''$ remains.

\subsection{Optimality conditions}

Minimising $\varepsilon-\muB n-\mue\,(n_e+n_\mu-\sum_XQ_Xn_X)$ with multipliers
$\lambda_q(\mathbf{q})\ge0$ for the constraints~\eqref{eq:pauli} gives, for every
species and momentum, with $\omega_X$ of Eq.~\eqref{eq:fx},
\begin{equation}
\begin{aligned}
  &\omega_X(k)
  \begin{cases} \ge0, & f_X=0,\\ =0, & 0<f_X<1,\\ \le0, & f_X=1,\end{cases}\\
  &\Phi_X(k) = \Nc^3\sum_q B_q^X\,\Phat_q(k),\\
  &\Phat_q(k) = \frac{(2\pi)^3}{2}\int\! d^3q\;\lambda_q(\mathbf{q})\,
     K\!\left(\left|\mathbf{q}-\mathbf{k}/\Nc\right|\right),
\end{aligned}
  \label{eq:phi}
\end{equation}
with $\muB=\mu_n$. The first case is Eq.~\eqref{eq:gen}. With the entropy term the three
cases become Eq.~\eqref{eq:fx}. At fixed densities the kinetic energy is linear in the occupation
probabilities and the constraints are linear. These conditions are then necessary and sufficient
for the minimum.

\subsection{Neutron matter above $\nq$}

Let $\Phi_n=18\Phat_d$. Neutrons with $k<\kbu$ have $0<f_n<1$, so $\omega_n=0$ and
$\Phi_n(k)=\mukin-E_N(k)$. This is the first line of Eq.~\eqref{eq:phin}.

The kernel obeys $(-\nabla_p^2+\Ls^{-2})K(p)=\delta^3(\mathbf{p})/(\Nc^3\Ls^2)$. Applied to
Eq.~\eqref{eq:phi} in the variable $\mathbf{p}=\mathbf{k}/\Nc$,
\begin{equation}
  (-\nabla_p^2+\Ls^{-2})\,\Phi_n(\mathbf{p}) \propto \lambda_d(\mathbf{p}),
  \label{eq:yuk}
\end{equation}
with a positive factor. The quark occupation $f_d$ obeys the same equation with the source
$18f_n/\Ls^2$. For $q<\kbu/\Nc$ the source is $1/\Ls^2$, and the regular solution is the constant 1
plus a term in $\sinh(q/\Ls)/q$. That term vanishes by the condition $f_d(0)=1$. Hence $f_d=1$ in
this range, and the constraint is active there. In the shell, with $q_{\rm bu}=\kbu/\Nc$ and
$s=q-q_{\rm bu}$, the solution that joins smoothly at $q_{\rm bu}$ is
\begin{equation}
  f_d(q) = 18-\frac{17}{q}\left[q_{\rm bu}\cosh\frac{s}{\Ls}+\Ls\sinh\frac{s}{\Ls}\right].
  \label{eq:fdshell}
\end{equation}
Since $\cosh y\ge1$ and $\sinh y\ge y$, $f_d<1$ for $s>0$. Above the shell $f_d$ falls as
$e^{-q/\Ls}/q$. The constraint is therefore inactive for $q>q_{\rm bu}$ and $\lambda_d=0$ there.
The solution of Eq.~\eqref{eq:yuk} that vanishes at infinity is then proportional to
$e^{-k/(\Nc\Ls)}/k$, and continuity at $\kbu$ gives the second line of Eq.~\eqref{eq:phin}.

For $k\ge\kbu$, Eq.~\eqref{eq:phin} gives
\begin{equation}
\begin{aligned}
  \omega_n(k) &= E_N(k)+\Phi_n(k)-\mukin = \frac{F(\kbu)-F(k)}{k\,e^{x}},\\
  F(k) &= k\,e^{x}\left[\mukin-E_N(k)\right],\qquad x=\frac{k}{\Nc\Ls}.
\end{aligned}
  \label{eq:omn}
\end{equation}
The derivative $F'$ vanishes where $\mukin=E_N+k^2/[E_N(1+x)]$. The right side increases with $k$,
so $F$ has a single maximum. From Eq.~\eqref{eq:kbu},
$\tfrac{17}{18}r\kbu=\ksh\,e^{-(x_{\rm sh}-x_{\rm bu})}$. With $\alpha=\tfrac{17}{18}r\kbu^2$, Eq.~\eqref{eq:mukin}
gives $\mukin-E_N(\kbu)=[E_N(\ksh)-E_N(\kbu)]\ksh^2/(\ksh^2-\alpha)$ and
$\mukin-E_N(\ksh)=[E_N(\ksh)-E_N(\kbu)]\alpha/(\ksh^2-\alpha)$. Hence $F(\ksh)=F(\kbu)$ and
$\omega_n(\ksh)=0$. The maximum of $F$ therefore lies between $\kbu$ and $\ksh$. It follows that
$\omega_n<0$ in the shell, $\omega_n>0$ above it, and $F'(\kbu)>0$.

The multiplier is positive. For $q<\kbu/\Nc$, Eq.~\eqref{eq:yuk} gives $\lambda_d\propto\nabla_p^2E_N
+(\mukin-E_N)/\Ls^2>0$. At the surface the derivative of $\Phi_n$ jumps and $\lambda_d$ has a
$\delta$-function of weight proportional to $\Phi_n(\kbu)[1/\kbu+1/(\Nc\Ls)]-\kbu/E_N(\kbu)$. This
equals $F'(\kbu)\,e^{-x_{\rm bu}}/\kbu$ and is positive.

All conditions of Eq.~\eqref{eq:phi} therefore hold at every density above $\nq$. The neutron
occupation of FKM is the minimiser in pure neutron matter. Since $\lambda_d$ is confined to
$q\le q_{\rm bu}$ and $\omega_n(\ksh)=0$ fixes $\mukin$, the multiplier is unique. The linear
program confirms these results for the three values of $\Ls$ (Sec.~\ref{sec:thrL}).

\subsection{Cost of a hyperon}

In neutron matter $\Phat_u=\Phat_s=0$, and for a species with fraction $B_d^X$ of $d$ quarks
Eq.~\eqref{eq:phi} gives $\Phi_X=\tfrac32B_d^X\Phi_n$. For the $\Lambda$, $\Phi_\Lambda=\Phi_n/2$,
and with Eq.~\eqref{eq:sp} at $n_\Lambda=0$ this is Eq.~\eqref{eq:omL}. The derivative of the
bracket of Eq.~\eqref{eq:cond} for $k<\kbu$ is $k(1/2E_N-1/E_\Lambda)<0$, so the maximum over this
range is at $k=0$. Below $\nq$, $\Phi_n=0$ and the cost is $E_\Lambda(k)+\UL-E_F-\UN$, lowest at
$k=0$. A positive $\omega_\Lambda$ for all $k$ means that the energy rises at first order in
$n_\Lambda$, and $n_\Lambda=0$ is a local minimum. Since $W$ and Eq.~\eqref{eq:eint} are not convex
in $n_\Lambda$, this does not exclude a distant lower minimum.

\subsection{First-order transitions}

Where the minimisation at fixed density gives a pressure that decreases with
density, the stable state at given $\muB$ is the solution with the largest
pressure. The EoS then follows the upper envelope of $P(\muB)$, and the density
jumps at the transition. Charge neutrality is imposed in each phase separately.
This is a Maxwell construction, appropriate when the surface tension between the
phases is large. For neutrons and $\Lambda$ hyperons alone there is a single conserved charge, and
the construction is exact. With charged particles a Gibbs construction with global charge neutrality could change
the transition interval, the density jump, the compositions and therefore the reduction
of maximum mass. It is not considered here.

\section{Numerical methods}
\label{sec:app}

\emph{Global minimisation.} The baryon momentum is discretised in 220 intervals up to
$1200\MeV$ and the quark momentum in 90 intervals up to $450\MeV$. The kinetic energy density is
computed with the HiGHS linear-programming code for $n$ from $0.25$ to $10\,\nz$, in steps of
$0.125\,\nz$ above $\nz$, and for $x_\Lambda$ from 0 to 1 in steps of $0.025$. Pressures follow by
differentiation at the minimum over $x_\Lambda$. The maximum masses without hyperons are $2.123$,
$1.981$ and $2.081\,\Msun$, against $2.120$, $1.978$ and $2.077\,\Msun$ from the closed form. An
independent construction of the convex hull of $\varepsilon(n)$ on interpolated tables changes the
reductions by at most $0.005\,\Msun$.

\emph{Iteration.} The baryon momentum is discretised in 240 intervals up to $2800\MeV$ and the
quark momentum in 28 intervals up to $1600\MeV$. The angular average of the kernel is computed
numerically. The discretised sum is normalised to reproduce Eq.~\eqref{eq:G} for a filled neutron
sphere. At fixed baryon density the occupations~\eqref{eq:fx} are computed for given multipliers
and potentials. The chemical potentials $\muB$ and $\mue$ are solved for the baryon density and
charge neutrality. The multipliers are then updated by $\lambda_q\to\max[0,\lambda_q+\eta(f_q-1)]$
with $\eta=T/2$, and the densities that enter the potentials are relaxed towards the new values.
The temperature is lowered in the steps $8$, $4$, $2$ and $1\MeV$. The potentials are numerical
derivatives of Eq.~\eqref{eq:eint}. The pressure is taken as $P=\muB n-\varepsilon$, without the
entropy term. A cold start uses matter without hyperons. A hot start begins from
matter rich in hyperons.

\emph{Stars.} The EoS is tabulated from $0.2$ to $9\,\nz$ in steps of $0.25\,\nz$
above $\nz$; for the model without interaction energy from $1$ to $16\,\nz$ in steps
of $0.5\,\nz$. The Tolman--Oppenheimer--Volkoff equations are integrated with a
fourth-order Runge--Kutta method and a radial step of $25$~m. No crust is attached.
The stars with and without hyperons are computed with the same program, grids and
temperature. The squared sound speed $dP/d\varepsilon$ on the stable branch, on the density
grid, is between 0 and 1 up to the central density for every EoS of Table~\ref{tab:octet}; its
largest value is $0.47$. For the $\Lambda$ alone with the iterative method and $\xi_\Lambda\ge0.4$ it
is at most $0.62$. At smaller $\xi_\Lambda$ some iterative solutions have a squared sound speed
above 1. For the mean-field potentials of Sec.~\ref{sec:hadronic} with the global minimisation it is at most $0.46$.

\emph{Thresholds.} The thresholds of Secs.~\ref{sec:thr} and \ref{sec:req} follow from
Eq.~\eqref{eq:cond} at zero temperature. Those of Table~\ref{tab:charged} follow from the
multipliers of the linear program. The first hyperons of Table~\ref{tab:octet} are read from the
iteration on the density grid, as the lowest density with a fraction above $10^{-3}$.

\emph{Tests.} The linear program reproduces the closed forms as stated in Sec.~\ref{sec:thrL}.
The iteration and the global minimisation agree for the $\Lambda$ alone. Their reductions differ by
at most $0.014\,\Msun$ for $\xi_\Lambda$ from $0.2$ to 1 in model~A and for $\xi_\Lambda=0.2$ in
models~B and C, and by at most $0.001\,\Msun$ for larger $\xi_\Lambda$ in models~B and C. At
$\xi_\Lambda=0$ they differ by up to $0.27\,\Msun$. There the iteration can stop at a solution of
higher energy, which is why Table~\ref{tab:loss} uses the global minimisation.
The iteration on its momentum grid gives lower nucleonic maximum masses, $2.111$, $1.954$ and
$2.053\,\Msun$ for models~A, B and C in neutron matter. This comes mainly from the coarse quark
grid. A grid of
$480\times112$ gives $2.120\,\Msun$ in model~A. The effect is common to the stars with and without hyperons. At $480\times112$, with a cold start only,
model~A with $\xi_\Lambda=2/3$ falls by $0.05$ for $c_\Lambda=0$ and $0.24\,\Msun$ for the lower edge of the band.
With $\xi_\Lambda=1$ it falls by $0.03$ and $0.15\,\Msun$. In $\beta$ equilibrium a linear program for
neutrons, protons and leptons gives $2.12$, $1.98$ and $2.07\,\Msun$, against $2.11$, $1.97$ and
$2.05\,\Msun$ from the iteration.
The reductions at $T=2$ and $1\MeV$ differ by at most $0.003\,\Msun$ for the $\Lambda$ alone and by
at most $0.009\,\Msun$ for the octet. For the octet in model~A the step to $0.5\MeV$ changes the
reductions of Table~\ref{tab:octet} by at most $0.003\,\Msun$.
For $\xi_Y\ge0.4$ the constraints~\eqref{eq:pauli} are violated by less than $0.4\%$ at the quark grid points at
densities inside the stars. Between the grid points $f_d$ reaches $1.006$ in models~A and B and
$1.018$ in model~C. For the $\Lambda$ alone with $\xi_\Lambda\ge0.4$ cold and hot starts can reach
different solutions at the same density. The solutions of both runs then enter the envelope of
$P(\muB)$ described in Appendix~\ref{sec:appmath}. With the $\Lambda$ as the only hyperon and without
protons, the octet program reproduces the program for neutrons and $\Lambda$ hyperons.

\bibliographystyle{apsrev4-2}
\bibliography{refs}

\end{document}